\documentclass[prl,showpacs,floatfix,twocolumn]{revtex4}

\usepackage{bm}
\usepackage{graphics}
\usepackage{graphicx}
\usepackage[dvips,usenames]{color}

\newcommand{\vs}{\vspace{0mm}}

\newcommand{\be}{\begin{equation}}
\newcommand{\ee}{\end{equation}}
\newcommand{\ba}{\begin{eqnarray}}
\newcommand{\ea}{\end{eqnarray}}

\newcommand{\AuthorTeam}{
\author{T.~S.~Bir\'o\footnote{On leave from KFKI RMKI Budapest, Hungary}} 
\affiliation{
 Institute for Theoretical Physics, University of Giessen,
 D 35392 Heinrich-Buff-Ring 16, Giessen, Germany
}
\author{A.~Peshier} 
\affiliation{
 Institute for Theoretical Physics, University of Giessen,
 D 35392 Heinrich-Buff-Ring 16, Giessen, Germany
}
}

\begin{document}

\title{{ Limiting temperature from a parton gas with power-law tailed distribution}}

\AuthorTeam

\date{{\bf June 21, 2005}}

\vs
\begin{abstract}

We combine Tsallis distributed massless partons to an effective thermal
prehadron spectrum by folding. A limiting temperature and a mass spectrum
combined of three exponentials emerge by this procedure. Meson and baryon
resonance spectra have different polynomial prefactors.
 
\end{abstract}

\maketitle

\vs

\vs
The idea to treat the bulk of mesonic and baryonic resonances as a statistical
system stems from Rolf Hagedorn\cite{HAGEDORN}. The mass spectrum is 
exponential, multiplied with an originally negative power of the mass, which
fact gives rise to a maximal, limiting temperature, $T_H$, when heating a
hadron gas. The exponential factor with the main part of the hadron
energy, i.e. the rest mass, is the most essential feature in this assumption,
power factors and depending on them the quantitative fit to the Hagedorn 
temperature, $T_H$, vary. 
A recent compilation of hadron resonances by Broniowski, Florkowski
and Glozman shows that this idea works
well beyond the data used for original fits, although baryons and mesons seem
to follow a slightly different line\cite{RESONANCES}. Explanations for the
origin of this exponential mass spectrum date back to the MIT bag model,
where this behavior was demonstrated by Kapusta\cite{BAG-LIMIT}.
Bugaev and others point out that such a system is a perfect
thermostat forcing the same temperature to any finite system in thermal
contact\cite{BUGAEV}. 
The picture of Hagedorn resonances also fit well to lattice QCD
results on the equation of state\cite{REDLICH} and is under consideration
in recent microscopic models of quark matter rehadronization in heavy-ion
collisions\cite{GREINER}.

\vs
In this paper we demonstrate how another approach may support the occurence
of an essentially exponential mass spectrum of hadrons. We consider massless
thermal partons, but with a generalized equilibrium distribution with
power-law tail asymptotics. This distribution will then be folded to mesonic
or baryonic one-particle distributions of energy, or - at zero rapidity -
transverse mass and transverse momenta, respectively.
Such stationary distributions are conform with
the basic laws of statistical physics and may be considered as a description
of the intermediate ($p_T\approx 1 - 4 $ GeV) part of the observed hadron 
spectra\cite{TSB-RHIC-SCHOOL,T-TEAM-FITS}. This approach delivers qualitatively
interesting results, in particular a characteristic difference between the 
mesonic and baryonic mass spectrum.

\vs
There are numerous occurrences of power-law tailed statistical distributions
in nature.
In particular hadron transverse spectra stemming from elementary particle or heavy ion 
collisions can be well fitted at mid rapidity by a formula reflecting 
$m_T$-scaling\cite{PHENIX,STAR,ZEUS,MAREK,BECK,TASSO,MT-SCALING,T-TEAM-FITS}: 
\be
 f(p_T) \sim \left(1 + m_T/E_c \right)^{-v}.
 \label{TSALLIS-DISTR}
\ee
Interpreting these spectra  in terms of the single particle energy, one considers
$E=m_T=\sqrt{p_T^2+m^2}$ for a relativistic particle with mass $m$.
This formula describes a Tsallis distribution, which was
conjectured earlier by using axiomatic thermodynamical arguments \cite{TSALLIS-ENTROPY}. 
This differs from the traditional interpretation of such spectral tails in
particle physics, when these are treated as non-equilibrium phenomena.
The very high-$p_T$ part, expected to stem from jet fragmentation, 
may still allow for a statistical interpretation below $p_T=6-8$ GeV.

\vs
Distributions with a power-law like tail are encountered
in several different statistical 
models\cite{MULT-NOISE,WILK,BIALAS,FLORKOW,KODAMA,RAF-BOLTZMANN-EQ,WAL-RAFELSKI,TSB-RHIC-SCHOOL,ASTRO}. 
They are investigated as generic distributions in the 
non-extensive thermodynamics\cite{TSALLIS-RULES,TSALLIS-WANG,TSALLIS-FOKKER}, 
based on a generalization of the Boltzmann-entropy, 
encountered first in informatics problems\cite{TSALLIS-ENTROPY,OTHER-ENTROPIES}. 
Without being able to exclude the non-equilibrium interpretation 
of the power-law tail in particle physics, 
we explore here some consequences of a stationary state with
Tsallis distributed extreme relativistic particles (massless partons).



\vs
We consider a massless parton gas with binary collisions
obeying a general, non-extensive energy composition rule\cite{NEBE},
$E_{12}=h(E_1,E_2)$. Whenever this rule is associative, the one-particle energy
can be mapped onto an additive quasi-energy $X(E_{12})= X(E_1)+X(E_2)$, with
the help of a strict monotonic function, obeying $X(0)=0$. For the Tsallis
stationary distribution one uses
\be
 X(E) = \frac{1}{a} \ln \left(1 + aE \right),
 \label{DEF-TSALLIS}
\ee
with an energy scale $a=1/E_c$ related to the microscopical dynamics.
This gives rise to the following non-extensive composition rule:
\be
 h(x,y) = x + y + a xy.
\ee
The stationary one-particle distribution under these conditions becomes
\be
 f(E_i) = \frac{1}{Z_1(\beta)} \exp\left(- X(E_i)/T \right),
 \label{ONE-DISTR}
\ee
with a temperature $T$ determined by the conserved total (quasi) energy
$X(E_{tot})$ and particle number. 
The one-particle partition function here is given by the phase space integral 
$ Z_1(\beta) \: = \: \int d\Gamma_j f(E_j)$.

\vs
In this paper we are interested in the energy distribution of large subsystems,
i.e. in the microcanonical distribution of the energy of $N$ particles
each following a Tsallis distribution given by eq.(\ref{ONE-DISTR}).
The general formula for the $N$-particle energy distribution, assuming
a factorization of the one-particle distributions is given by
\be
 F_N(E) = \Delta E  X'(E) \int \prod_{i=1}^N d\Gamma_i \:
 \delta(X(E)-\sum_{i=1}^N X(E_i)) \, f(E_i).
 \label{MANY-DISTR}
\ee
Here $X'(E)$ stands for the derivative of the strict monotonic mapping function
$X(E)$, $\Delta E$ is the width of the $N$-particle energy shell and the 
$d\Gamma_i$ integration measures refer to the one-particle phase space factors.
The factorization is usually a good approximation for values of $N$
being still negligible besides the total number of particles.
A check of this formula for $N=1$ expresses the one-particle energy distribution
as being proportional to the one-particle phase space distribution:
\be
 F_1(E) \: = \: \frac{V}{2\pi^2} E^2 \Delta E \, f(E).
\ee
Since $F_N(E)$ contains an $X'(E)$ factor,
$F_N(E) dE = g_N(X) dX$ relates it to the distribution
of the $N-$particle quasi energy, $X=X(E)$. 
As it was shown in Ref.\cite{NEBE}, in certain kinetic models leading to
a stationary state of the non-extensive thermodynamics, $X(E)=\sum_i X(E_i)$ is the
conserved quantity and the quasi-energy can be regarded as the physical energy
of composite $N-$particle systems.

\begin{figure}
\begin{center}
 \includegraphics[width=0.30\textwidth,angle=-90]{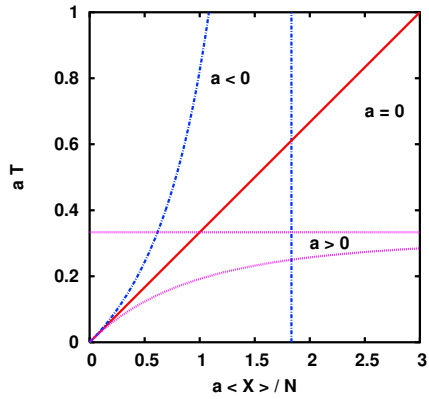}
 \end{center}
 \caption{ \label{LIMIT_TEMP} 
 The temperature, $T$, as a function of the quasi-energy per particle,
 $\langle X \rangle / N$, in units of the Tsallis energy scale $E_c=1/a$.
 The linear relation $E/N=3T$ known for massless particles is realized
 for $a=0$. For repulsive corrections, $a>0$, a limiting temperature
 occurs, $T_H=1/3a=E_c/3$.
 }
 \end{figure}

\vs
Using the Fourier-expansion of
the constraint on the sum of the quasi-energies, $X_i=X(E_i)$, the expression
(\ref{MANY-DISTR}) factorizes:
\be
 g_N(X) \: = \: \Delta E \int_{-\infty}^{+\infty} \frac{ds}{2\pi} \, e^{isX} \,
 \prod_{j=1}^N \left[ \int d\Gamma_j f(E_j) e^{-isX(E_j)}   \right].
 \label{MANY-FACTOR}
\ee
Utilizing now the equilibrium one-particle quasi-energy distribution,
(\ref{ONE-DISTR})
we obtain
\be
 \int d\Gamma_j f(E_j) e^{-isX(E_j)} \: = \: \frac{Z_1(\beta+is)}{Z_1(\beta)}.
 \label{Z-RATIO}
\ee
The $N$-particle quasi-energy distribution we are seeking for is then given 
in a form normalized to one in an energy shell of width $\Delta E$ as
\be
 g_N(X) = \Delta E \int_{-\infty}^{+\infty} \! \frac{ds}{2\pi} \, e^{isX} 
  \, \left(\frac{Z_1(\beta+is)}{Z_1(\beta)} \right)^N.
  \label{BIG-INTEGRAL}
\ee
Such integrals may be evaluated in a saddle point approximation,
which is good for large $N$ as long as no singularity has been encountered
in the expansion of $\ln Z_1(\beta+is)$. The result is a Gaussian
distribution in $X(E)$. 
Irrespective to this approximation, as long as the factorization assumption
is valid, the exact expectation value is given by
\be
 \langle X(E) \rangle \: = \: -N \frac{\partial}{\partial \beta} \ln Z_1(\beta),
 \label{EXPECTO-X}
\ee
and the square width by
\be
 \delta X(E)^2 \: = \: N \frac{\partial^2}{\partial \beta^2} \ln Z_1(\beta).
 \label{WIDTH-X}
\ee
In case of the Tsallis distribution using (\ref{DEF-TSALLIS}) for massless
particles one obtains
\be
 Z_1(\beta) = \frac{V_d}{(2\pi)^d} (d-1)! \, \prod_{k=1}^d B_k^{-1}
 \label{Z-TSALLIS}
\ee
with spatial volume $V_d$ in $d$ dimensions and with the factors $B_k=\beta-ka$. 
The expectation value of the quasi-energy per particle becomes
\be
 \epsilon_1 = \frac{\langle X(E) \rangle }{N} \: = \: \sum_{k=1}^d B_k^{-1},
\label{ENER-PER-PART}
\ee
while the unit square width contribution is given by
\be
 \delta_1 = \frac{\delta X^2}{N} \: = \: \sum_{k=1} B_k^{-2}.
 \label{WIDTH-PER-PART}
\ee
All these expressions (\ref{Z-TSALLIS},\ref{ENER-PER-PART},\ref{WIDTH-PER-PART})
loose their conventional interpretability for $\beta \le da$ in $d$ dimensions.
The value, $T_H=1/(da)$ is a limiting temperature for positive values
of the parameter $a$, at which the physically relevant quasi-energy per
particle diverges, so there is no use of further heating at this temperature.
More and more energy given to the system would raise the temperature less and less.
For $a<0$, i.e. for attractive correction, the energy
per particle is limited by $(1+1/2+1/3)E_c$  from above, but any
temperature may occur (cf. Fig.\ref{LIMIT_TEMP}).

\vs


\vs
The $N$-particle energy distribution can be obtained also exactly in this case.
The Fourier-integral (\ref{BIG-INTEGRAL}) has $N$-fold poles at the values
$s_k=-iB_k$ for each directional degree of freedom $k=1,\ldots,d$. The evaluation
of such integrals is somewhat involved in a general number of dimensions, so
we restrict our further analysis to the cases $d=1$ and $d=3$.
For a one-dimensional Tsallis distribution we obtain
\be
 g_{N+1}(X) \: = \: \Delta E  B_1 \frac{(B_1X)^N}{N!} \, 
 	\exp\left( X/E_c - X/T  \right).
\label{ONE-DIM}
\ee
In this case a Hagedorn exponential emerges with the limiting
temperature $T_H=E_c$. Considering on the other hand $F_{N+1}(E)=g_{N+1}(X) X'(E)$,
i.e. the $N+1$-particle distribution of the (not conserved) naive
energy expression only, the exponetially growing factor does not occur.
It is due to $X'(E)=e^{-aX}$ for the Tsallis distribution:
\be
 F_{N+1}(E) \: = \:  \Delta E 
 \frac{{B_1^{N+1}}{\left(\ln(1+aE)\right)}^N}{a^N \, N!}
 (1+aE)^{-1/aT}.
\ee
However, still $\beta>a$ or $T<T_H$ must be satisfied, and for positive $a$ values
increasing quasi-energy per particle does not raise the temperature above $T_H$.

\vs
The corresponding expression in $d=3$ dimensions is more involved.
There occur three different exponential factors with lowest limiting
temperature $T_H=E_c/3$. The general dependence on $\beta$ can be factorized
out by shifting the integration variable $s$ to $s+i\beta$:
\be
 g_{N+1}(X) \: = \: \frac{ V^{N+1}\, \Delta E}{(\pi^2Z_1(\beta))^{N+1}}
 e^{-\beta X} \, \Phi_N(X),
\label{THREE-DIM}
\ee
with $\Phi_N(X)$ depending on the dynamical input parameters only, but not
on the temperature $T=1/\beta$:
\be
 \Phi_N(X) \: = \: \int\frac{ds}{2\pi} e^{isX} 
 	\left(\prod_{k=1}^3 B_k^{-1}(is)\right)^{N+1}.
\ee
Factorizing out the ideal thermal factor $e^{-\beta X}/Z_N$, the rest can be
regarded as the mass spectrum of the $N$-parton system at $X=m$ and $\Delta E =
\Delta m$:
\be
  \rho_N(m) \: = \: \left(\frac{V}{\pi^2}\right)^{N} \, \Phi_{N-1}(m).
\label{MASS-SPECTRUM}
\ee
The functions $\Phi_N(X)$ obey the recursion rule
\be
 \Phi_{N}(X) = \int_0^X\!\!dt \, \Phi_{N-1}(t) \Phi_0(X-t)
\ee
with the starting point of the recursion,
\be
 \Phi_0(X) = \frac{1}{2a^2} \left(e^{3aX}-2e^{2aX}+e^{aX} \right).
\label{START}
\ee
Particular important cases are $N=0$ (partons), $N=1$ (mesons or diquarks)
and $N=2$ (direct baryons). Besides the already known $\Phi_0(X)$
(eq. \ref{START}) we obtain
\be
 \Phi_1(X) = \frac{(aX-3)e^{3aX}+4aXe^{2aX}+(aX+3)e^{aX}}{4a^5} 
\ee
and
\ba
 \Phi_2(X)
 &=&
 \frac{1}{16a^8} \left[
  \left( (aX)^2-9aX+24 \right)e^{3aX}
 \right.
 \nonumber \\
 &&
 \qquad\quad
 \left.
  -8\left( (aX)^2+6 \right)e^{2aX}
 \right.
 \nonumber \\
 &&
 \qquad\quad
 \left.
  +\left( (aX)^2+9aX+24 \right)e^{aX}
 \right]
\ea
For any $N$ the result contains three exponentials giving rise to a
lowest limiting temperature of $T_H=E_c/3=1/3a$ for positive values of the
parameter $a$. 

\begin{figure}
\vspace*{-25mm}
\begin{center}
 \includegraphics[width=0.40\textwidth,angle=-90]{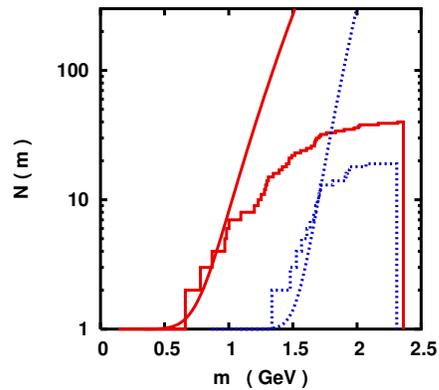}
 \end{center}
 \caption{ \label{MASS_SPECTRUM} 
   The integrated, non-degenerated mass spectrum of 
   mesonic and baryonic resonances from the
   2004 compilation by the Particle Data Group\cite{PDG} is compared with the
   respective cumulative numbers of two and three-parton clusters.
 }
 \end{figure}

\vs
Fig.\ref{MASS_SPECTRUM} shows the non-degenerate, integrated mass spectra.
Values published on the Particle Data Group
homepage are summed up in mass histograms. The respective numbers
are fitted as
$N_M=1+Af_1((m-m_M)/3T_H)$ and $N_B=1+A^2f_2((m-m_B)/3T_H)$ 
with $A=V/a^3\pi^2$ and the integral functions
$f_n(x)=\int_0^x \Phi_n(t) dt$. 	
The fits to the data are most sensitive to the value of $T_H$,
which however may be compensated by changing the assumed volume, $V$.
Keeping $m_M=0.14$ GeV and $m_B=0.94$ GeV, only $A$ and $T_H$
are varied. Our best fit gives a relatively high value, $T_H=0.35$
GeV (meaning $E_c=1.05$ GeV) and a volume of $V=261$ $fm^3$
(a sphere with a radius of $4$ fm, or a box with a length of $6.4$ fm).
Above the masses where the data seem to deviate from the fast growing part,
the fit cannot be followed any more. According to ref.\cite{RESONANCES},
newest data raise the experimental curve higher.
Our idea, different from both the string and bag model consideration,
seem to agree with the difference between the meson and baryon mass
spectra, as well as with a polynomial upcurving of the baryon spectrum.

\vs


\vspace{3mm}
\vs

In conclusion we pointed out that Tsallis distributed massless partons
can be combined to an effective mesonic and baryonic mass spectrum by
considering the conserved quasi-energy as the hadron energy. Besides
an ideal thermal factor, $e^{-\beta X}/Z_N $, a further energy dependent
factor results from the folding of parton distributions. It can be
regarded as a thermal (pre)hadron mass
spectrum emerging from a statistical hadronization picture.
The prediction of this folding, while having two parameters (a volume
and the Hagedorn temperature), gives an acceptable qualitative agreement
with the known hadronic mass spectrum. In this picture a natural difference
emerges between mesonic and baryonic resonances due to their different
foldness by parton coalescence. The characteristic temperature,
$T_H=1/3a \approx 350$ MeV is a limiting temperature: one cannot
increase the temperature above this value, not even with an infinite
amount of energy.
The parameter $E_c=1/a=3T_H \approx 1.05$ GeV provides at the same time the scale where
the power-law tail of individual $p_T$-spectra starts to dominate the
exponential part, and it is intimately related to the typical pair-interaction
energy, due to \hbox{$h(E_1,E_2)-E_1-E_2=E_1E_2/E_c$.}

\vspace{3mm}
\vs
{\bf Acknowledgment}

This work has been supported by BMBF, by the Hungarian National Science Fund OTKA
(T49466) and by DFG due to a Mercator Professorship for T.S.B.




%

\end{document}